\documentclass[10pt,letterpaper,twocolumn]{article} 

\usepackage{ol2}
\usepackage[draft]{hyperref}
\usepackage{amssymb}
\newcommand{\fref}[1]{Fig.~\ref{fig:#1}}
\newcommand{\rthz}[1]{\mathrm{#1}/\sqrt{\mathrm{Hz}}}

\begin{document}
\twocolumn[ 


\title{Control and tuning of a suspended Fabry-Perot cavity using
  digitally-enhanced heterodyne interferometry}

\author{John Miller,$^{*}$ Silvie Ngo, Adam J.\ Mullavey, Bram J.\ J.\
  Slagmolen,\\Daniel A.\ Shaddock and David~E.~M$^\mathrm{c}$Clelland}

\address{Centre for Gravitational Physics, The Australian
  National University,\\
  Canberra, ACT, 0200, AUSTRALIA}

\email{*john.miller@anu.edu.au}

\begin{abstract}
  We present the first demonstration of real-time closed-loop control
  and deterministic tuning of an independently suspended Fabry-Perot
  optical cavity using digitally-enhanced heterodyne interferometry,
  realising a peak sensitivity of $\sim$10~$\rthz{pm}$ over the
  10--1000~Hz frequency band. The methods presented are readily
  extensible to multiple coupled cavities. As such, we anticipate that
  refinements of this technique may find application in future
  interferometric gravitational-wave detectors.
\end{abstract}


] 

Fabry-Perot optical resonators continue to play an important role as
frequency references and displacement sensors across a variety of
disciplines. Several established techniques exist for deriving
resonator sensing and control signals, the most prevalent being the
Pound-Drever-Hall (PDH) method \cite{Drever83}. However, all of these
techniques suffer from the same shortcoming -- useful signals are only
realised in a small neighbourhood around resonance. This
characteristic reduces sensing range and can make it difficult to
bring a cavity to resonance from an initially uncontrolled state, a
process known as \emph{lock acquisition}.

Lock acquisition proved a particularly challenging problem for the
first generation of gravitational-wave interferometers (GWIFOs) and,
due to increased optical complexity, the solutions developed
\cite{Evans2002,Acernese2008short} are not appropriate for
contemporary instruments (e.g.~\cite{Harry2010}). Hence, this issue
has been the subject of significant research interest in recent years,
resulting in the development of acquisition schemes for second
generation GWIFOs (e.g.~\cite{WardThesisShort}). More recently, tools
have been developed to supplement these schemes in an effort to make
the lock acquisition process more deterministic \cite{Mullavey2012,
  Izumi2012}. However, these tools are limited to a single degree of
freedom and may not be immediately compatible with proposed
third-generation detectors (e.g.~\cite{Punturo2010short}). For these
reasons, readouts based on the newly developed digitally-enhanced
heterodyne interferometry (DEHI) technique \cite{Shaddock07} are of
great interest to the gravitational-wave community.

DEHI augments the standard heterodyne interferometry technique with an
additional radio-frequency binary modulation-demodulation stage to
provide a number of additional features without compromising the
underlying heterodyne sensitivity. Specifically, it allows
high-dynamic-range multiplexed measurements to be performed whilst
simultaneously suppressing noise due to electronics and scattered
light.

The multiplexing capabilities of DEHI have been verified in fibre
strain measurements \cite{Wuchenich2011} and its noise performance has
been explored using a low-finesse fixed-mirror Fabry-Perot cavity
\cite{deVine2009}.  We build upon this work to present the first
demonstration of real-time control and tuning of an independently
suspended Fabry-Perot optical cavity using DEHI. These results are of
relevance to the problem of lock acquisition in future GWIFOs.

In order to measure the detuning of a Fabry-Perot optical cavity we
employ DEHI in transmission to perform heterodyne measurements on two
beams, one which passes directly through the cavity and another which
undergoes a round trip before being transmitted. Both measurements are
performed relative to the same reference beam so that the difference
between the resulting phases is the round-trip cavity phase.

\begin{figure}[htbp!]
\centerline{
\includegraphics[width=\columnwidth]{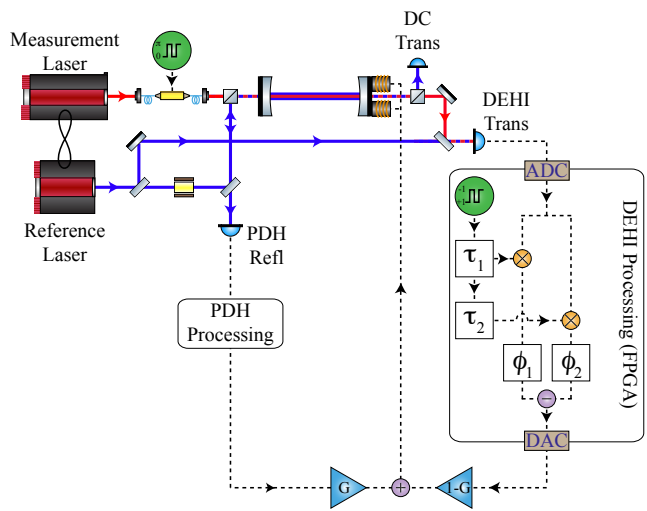}}
\caption{(Colour online) Schematic showing our cavity apparatus
  addressed by two measurement systems, DEHI and PDH. The DEHI
  reference laser doubles as the input source for PDH readout. Cavity
  length can be controlled by applying either PDH or DEHI signals to
  coil-magnet actuators on the resonator's end mirror.}
\label{fig:setup}
\end{figure}
A schematic of our experimental setup is shown in \fref{setup}. Our
cavity was formed between two `Tip-Tilt' suspension systems
\cite{Slagmolen2011}. It was 1.3~m long and had a finesse of
$\sim$300. The measurement laser was an Innolight Prometheus (1064~nm
output). This source was offset phase-locked to a Lightwave 126 laser
which provided both the DEHI reference beam and the input light for
corroboratory PDH measurements. All digital signal processing was
carried out using a commercial field-programmable gate array (Xilinx
Virtex-5 LX110).

The measurement beam was first phase modulated with a pseudo-random
noise (PRN) code of length \hbox{$2^{15}-1$}. The chipping frequency
was set to 115.3~MHz, matching the cavity's free spectral range. In
this way a signal which completes $n$ round trips through the cavity
is delayed by $n$ chips.

The beam transmitted through the cavity was combined with the
reference beam and detected by a wide-bandwidth photodetector (New
Focus 1811). The photodiode output was sampled at 115.3~MHz and
divided into two processing channels. The first channel was
demodulated with the PRN code delayed by $\tau_1$ to isolate the
straight-through beam. The second channel was demodulated with the PRN
code delayed by $\tau_1+\tau_2=\tau_1+nT_{\mathrm{PRN}}$, where
$T_{\mathrm{PRN}}$ is the PRN code chip-period, to isolate the
$n^{\mathrm{th}}$-round-trip beam.

The demodulated signals were subsequently passed to independent
digital phase meters. These phase meters continually modify their
local oscillator to match its frequency to that of the input signal,
integrating the frequency error to retrieve phase information. It is
this closed-loop operation which endows DEHI measurements with their
impressive dynamic range.

The outputs of the two phase meters were finally subtracted to yield a
direct measure of the round-trip cavity phase. This difference was
used as the error signal in a feedback loop controlling the cavity's
resonant state. Correction signals were applied to the cavity end
mirror using coil-magnet actuators.

\begin{figure}[htbp!]
\centerline{
\includegraphics[width=\columnwidth]{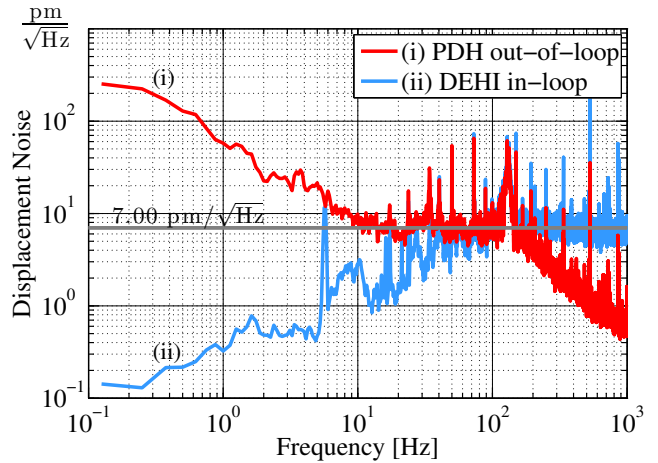}}
\caption{(Colour online) Cavity displacement noise measured via the
  Pound-Drever-Hall technique whilst the cavity was locked on
  resonance using digitally-enhanced heterodyne interferometry.}
\label{fig:results}
\end{figure}
Of utmost importance to any interferometric sensing technique is its
noise performance or sensitivity. This quantity was evaluated by
locking the cavity on resonance using the DEHI system and employing a
standard PDH sensing scheme as an out-of-loop sensor.

The results, presented in \fref{results}, show both the in-loop DEHI
error signal and the out-of-loop PDH error signal. Together, these
traces reveal that the DEHI scheme employed here is limited by white
sensing noise at the 7~$\rthz{pm}$ level with performance degrading
toward lower ($\lesssim$20~Hz) frequencies.

In a DEHI measurement signals arrive at the photodetector not only
from the desired straight-through and round-trip beams but also from
light which undergoes multiple round trips within the cavity. Since
DEHI provides only finite attenuation of unwanted signals, these
additional components limit sensitivity and give rise to the white
noise observed.

The broadband nature of the DEHI noise floor may limit usefulness in
noise-critical applications, especially in the context of multiply
suspended optics. For example, based on Simulink modelling of the
Advanced LIGO suspensions, we estimate that in order for DEHI to be
useful as a lock acquisition tool the white noise floor must be
reduced to $\sim$1~$\rthz{pm}$. This requirement is enforced to
prevent saturation of the suspension actuators. Fortunately, however,
the level of the DEHI noise floor scales as
$\sqrt{T_{\mathrm{PRN}}}$. This dependency allows one to leverage
future advances in digital signal processing to improve
performance. Based on currently available hardware we believe that
sub-picometre noise levels are realistically achievable.

Below $\sim$20~Hz, trace (i) in \fref{results} represents an upper
bound to the DEHI sensitivity.  Low-frequency noise is introduced by
the DEHI system but is associated with external factors. For instance,
alignment fluctuations affect DEHI more strongly than PDH as DEHI is a
non-resonant technique which does not benefit from the spatial mode
filtering of the cavity. A standard auto-alignment system would
substantially mitigate this problem. Acoustic couplings are also known
to adversely affect our bench-top apparatus. Indeed, the prominent
feature centred about 130~Hz is the acoustic resonance of the
intra-cavity beam tube introduced to ameliorate this low frequency
noise source. Vacuum operation would yield significant improvements in
this area.

The unity gain frequency of the DEHI loop was set to $\sim$100Hz,
where the free-swinging displacement noise intersected the DEHI noise
floor, so that injected noise would not compromise the pendulums'
isolation. This choice explains the observed high-frequency roll-off.

DEHI offers excellent sensitivity, compatible with the vast majority
of industrial and scientific measurements. Nevertheless, some extreme
endeavours require performance which surpasses that presented here. In
such cases DEHI can be used to acquire initial control before
deferring to a second sensor. Mindful of this, we detail in
\fref{handoff} the transfer of control over our independently
suspended optical cavity from the high-dynamic-range DEHI system to a
lower-noise PDH readout.

\begin{figure}[htbp!]
\centerline{
\includegraphics[width=\columnwidth]{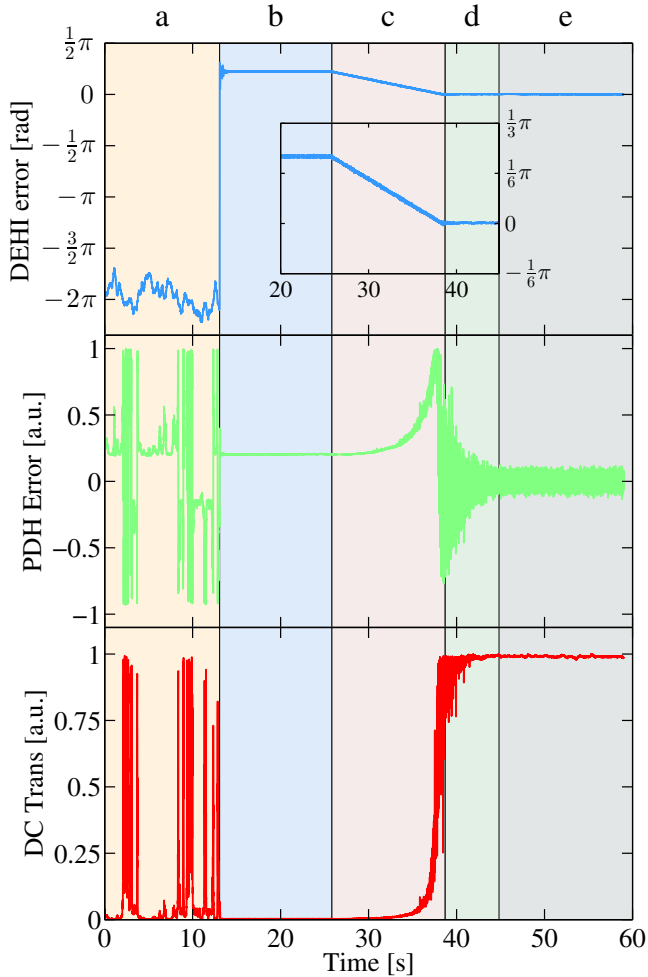}}
\caption{(Colour online) Automated acquisition of cavity length
  control using digitally-enhanced heterodyne interferometry. Top --
  DEHI error signal (i.e.~round-trip cavity phase). Middle --
  Normalised Pound-Drever-Hall error signal. Bottom -- Normalised
  reference laser cavity transmission. Data were sampled at
  $\sim$200~Hz. The division of the axes is discussed in the main
  text.}
\label{fig:handoff}
\end{figure}
Each stage of the transfer is now described according to the
alphabetic labels in \fref{handoff}. a)~Initially the cavity is freely
swinging, its motion driven by environmental disturbances. b)~The
cavity is then locked at a pre-determined offset from resonance, far
from any higher-order modes or sidebands, using the omnipresent DEHI
signal. c)~The offset is reduced in a controlled manner until the
cavity arrives within the linewidth of the desired carrier
resonance. d)~At this point the PDH signals become quasi-linear and
control over cavity length is transferred from the DEHI loop to the
PDH loop by simultaneously adjusting their respective gains (G in
\fref{setup}). e)~The cavity is now under sole control of the
PDH-based feedback loop. At all stages of the transfer control signals
were applied to the end mirror's coil magnet actuators.

These data confirm that DEHI has the capability to produce useful
sensing signals at any cavity detuning. This large dynamic range,
combined with previous demonstrations of multi-mirror multiplexing
\cite{Wuchenich2011}, engenders confidence that the technique will
perform as expected if applied to coupled-cavity interferometers.

Work is currently under way to extend the DEHI concept from
longitudinal to angular control. This would result in a wide-range
alignment sensor which does not have cavity resonance as a
prerequisite to its operation. Such a system may be capable of
replacing traditional optical levers. Techniques are also being
developed to allow DEHI to operate in reflection and with improved
sensitivity \cite{Sutton2012}.

\section*{Acknowledgements}
The authors thank Lisa Barsotti for valuable comments during the
preparation of this manuscript. This work was supported by the
Australian Research Council (ARC). JM is the recipient of an ARC Post
Doctoral Fellowship (DP110103472). This paper has been assigned LIGO
Laboratory document number LIGO-P1200100.


\pagebreak

\section*{Informational Fourth Page}


\begin{thebibliography}{10}
\newcommand{\enquote}[1]{``#1''}

\bibitem{Drever83}
R.~W.~P. {Drever}, J.~L. {Hall}, F.~V. {Kowalski}, J.~{Hough}, G.~M. {Ford},
  A.~J. {Munley}, and H.~{Ward}, \enquote{{Laser phase and frequency
  stabilization using an optical resonator},} Applied Physics B: Lasers and
  Optics \textbf{31}, 97--105 (1983).

\bibitem{Evans2002}
M.~{Evans}, N.~{Mavalvala}, P.~{Fritschel}, R.~{Bork}, B.~{Bhawal},
  R.~{Gustafson}, W.~{Kells}, M.~{Landry}, D.~{Sigg}, and R.~{Weiss},
  \enquote{{Lock acquisition of a gravitational-wave interferometer},} Optics
  Letters \textbf{27}, 598--600 (2002).

\bibitem{Acernese2008short}
{F. Acernese et al.}, \enquote{{Lock acquisition of the Virgo gravitational
  wave detector},} Astroparticle Physics \textbf{30}, 29--38 (2008).

\bibitem{Harry2010}
G.~M. {Harry} and {the LIGO Scientific Collaboration}, \enquote{{Advanced LIGO:
  the next generation of gravitational wave detectors},} Classical and Quantum
  Gravity \textbf{27}, 084006 (2010).

\bibitem{WardThesisShort}
{R.\ L.~Ward}, Ph.D. thesis, California Institute of Technology (2010).

\bibitem{Mullavey2012}
A.~J. Mullavey, B.~J.~J. Slagmolen, J.~Miller, M.~Evans, P.~Fritschel, D.~Sigg,
  S.~J. Waldman, D.~A. Shaddock, and D.~E. McClelland, \enquote{Arm-length
  stabilisation for interferometric gravitational-wave detectors using
  frequency-doubled auxiliary lasers,} Opt. Express \textbf{20}, 81--89 (2012).

\bibitem{Izumi2012}
K.~Izumi, K.~Arai, B.~Barr, J.~Betzwieser, A.~Brooks, K.~Dahl, S.~Doravari,
  J.~C. Driggers, W.~Z. Korth, H.~Miao, J.~Rollins, S.~Vass, D.~Yeaton-Massey,
  and R.~X. Adhikari, \enquote{Multicolor cavity metrology,} J. Opt. Soc. Am. A
  \textbf{29}, 2092--2103 (2012).

\bibitem{Punturo2010short}
{M. Punturo et al.}, \enquote{{The third generation of gravitational wave
  observatories and their science reach},} Classical and Quantum Gravity
  \textbf{27}, 084007 (2010).

\bibitem{Shaddock07}
D.~A. {Shaddock}, \enquote{{Digitally enhanced heterodyne interferometry},}
  Optics Letters \textbf{32}, 3355--3357 (2007).

\bibitem{Wuchenich2011}
D.~M.~R. Wuchenich, T.~T.-Y. Lam, J.~H. Chow, D.~E. McClelland, and D.~A.
  Shaddock, \enquote{Laser frequency noise immunity in multiplexed displacement
  sensing,} Opt. Lett. \textbf{36}, 672--674 (2011).

\bibitem{deVine2009}
G.~de~Vine, D.~S. Rabeling, B.~J.~J. Slagmolen, T.~T.-Y. Lam, S.~Chua, D.~M.
  Wuchenich, D.~E. McClelland, and D.~A. Shaddock, \enquote{Picometer level
  displacement metrology with digitally enhanced heterodyne interferometry,}
  Opt. Express \textbf{17}, 828--837 (2009).

\bibitem{Slagmolen2011}
B.~J.~J. Slagmolen, A.~J. Mullavey, J.~Miller, D.~E. McClelland, and
  P.~Fritschel, \enquote{Tip-tilt mirror suspension: Beam steering for advanced
  laser interferometer gravitational wave observatory sensing and control
  signals,} Review of Scientific Instruments \textbf{82}, 125108 (2011).

\bibitem{Sutton2012}
A.~J. Sutton, O.~Gerberding, G.~Heinzel, and D.~A. Shaddock, \enquote{Digitally
  enhanced homodyne interferometry,} Opt. Express \textbf{20}, 22195--22207
  (2012).

\end{thebibliography}

\end{document}